\documentclass[10pt,a4paper,twoside]{article}
\usepackage{epsfig}
\usepackage{baltlat6}
\usepackage{wrapfig}
\usepackage{longtable}
\usepackage{here}
\usepackage{array}
\begin{document}
\ \
\vspace{0.5mm}
\setcounter{page}{33}
\vspace{8mm}

\titlehead{Baltic Astronomy, vol.\,18, 33--52, 2009}

\titleb{PHOTOMETRY AND CLASSIFICATION OF STARS
AROUND\\ THE REFLECTION NEBULA NGC\,7023 IN CEPHEUS.\\ II. INTERSTELLAR
EXTINCTION AND CLOUD DISTANCES}

\begin{authorl}
\authorb{K.~Zdanavi\v{c}ius}{},
\authorb{J.~Zdanavi\v{c}ius}{},
\authorb{V.~Strai\v{z}ys}{} and
\authorb{M. Maskoliunas}{}
\end{authorl}

\moveright-3.2mm
\vbox{
\begin{addressl}
\addressb{}{Institute of Theoretical Physics and Astronomy, Vilnius
University,\\ Go\v{s}tauto 12, Vilnius, LT 01108, Lithuania}
\end{addressl}
}
\submitb{Received 2009 May 20; accepted 2009 June 24}

\begin{summary} Interstellar extinction is investigated in a 1.5 square
degree area in the direction of the reflection nebula NGC\,7023 at
$\ell$\,=\,104.1\degr, $b$\,=\,+14.2\degr.  The study is based on
photometric classification and the determination of interstellar
extinctions and distances of 480 stars down to $V$ = 16.5 mag from
photometry in the {\it Vilnius} seven-color system published in Paper I
(2008).  The investigated area is divided into five smaller subareas
with slightly different dependence of the extinction on distance.  The
distribution of reddened stars is in accordance with the presence of two
dust clouds at 282 pc and 715 pc, however in some directions the dust
distribution can be continuous or more clouds can be present.
\end{summary}


\resthead{Photometry and classification of stars around NGC\,7023. II.}
{K. Zdanavi\v{c}ius, J. Zdanavi\v{c}ius, V. Strai\v{z}ys, M.
Maskoliunas}

\sectionb{1}{INTRODUCTION}

The distances to star-forming regions in the Cepheus Flare, an
out-of-plane concentration of interstellar dust and molecular clouds,
are still unknown to sufficient accuracy, see the recent review by Kun
et al.  (2008).  This our investigation is an attempt to determine more
reliable distances and extinctions of dust clouds in the direction of
the reflection nebula NGC\,7023, illuminated by the young high-mass star
HD\,200775 and surrounded by the dust cloud TGU\,629 (Dobashi et al.
2005).

In our earlier paper (Zdanavi\v{c}ius et al. 2008, hereafter Paper I) we
determined in this area the magnitudes and color indices in the {\it
Vilnius} seven-color photometric system for 1240 stars down to
$V$\,$\approx$\,16.7 mag.  The published catalog for most of the stars
also contains two-dimensional spectral types determined by interstellar
reddening-free methods from the multicolor photometric data.

In the present paper we apply the classification results of Paper I for
determining the distribution of the interstellar dust with distance
in the 1.5 square degree area around the NGC\,7023 nebula, using only
the selected stars with the most reliable spectral types. In Section
2, we describe the classification methods based on the interstellar
reddening-free parameters, used for determining the spectral types, and
calculate interstellar reddenings and extinctions of the stars.  The
distribution of interstellar dust in the area is investigated in Section
3 and the discussion and summary of the results are given in Section 4.

\sectionb{2}{TWO-DIMENSIONAL PHOTOMETRIC CLASSIFICATION}

    For the classification of stars a few different codes using slightly
different spectral standards were used.

1. COMPAR code is based on the $\sigma Q$ method described by
Strai\v{z}ys et al.  (1992, 2002).  The method uses matching 14
different interstellar reddening-free $Q$-parameters of a program star
to those of about 8400 standard stars of various spectral and luminosity
classes, metallicities and peculiarity types.  The results of the
classification are spectral and luminosity classes and the indication of
peculiarity.  Several varieties of the code and sets of standards were
used.

2. xqKLAS code uses the xq-method described by Zdanavi\v cius (2005).
The method is based on a new concept of reddening-free parameters ($q$)
and a `virtual' quantity of the interstellar dust ($x$). 1418 standards
were formed by calculating the mean dereddened color indices for 89
spectral subclasses (in most cases, for each one subclass and for
late-type stars for each 0.25 subclass) and the 17 values of the
absolute magnitude, $M_V$.  The results of the classification are
spectral class and absolute magnitude.

3. TINKLAS code classifies stars using six $Q$,$Q$ diagrams
described in Strai\v{z}ys (1992) monograph.  Each of them is formed from
two reddening-free $Q$-parameters and calibrated in terms of spectral
classes and absolute magnitudes.  The results are spectral class and
absolute magnitude.

Spectral classes and absolute magnitudes of stars determined by the
methods (2) and (3) were used to estimate their luminosity classes
taking the calibration of MK spectral types in absolute magnitudes from
Strai\v{z}ys (1992).  Then the spectral and luminosity classes
determined by the three methods for each star were weighted and
averaged.  The intrinsic color indices used in determining the
interstellar extinction and the distance are also taken from
Strai\v{z}ys (1992).

As it was stated in Paper I, the $J$--$H$ and $H$--$K_s$ color indices
from the 2MASS survey (Cutri et al. 2003; Skrutskie et al. 2006) in
some cases were helpful for the identification of K and M dwarfs.

\sectionb{3}{INTERSTELLAR EXTINCTIONS AND DISTANCES}

The interstellar reddenings $E_{Y-V}$ of 480 stars with the most
reliable classification were determined as differences between the
observed color indices $Y$--$V$ given in Table 2 of Paper I and the
intrinsic color indices ($Y$--$V$)$_0$ for a given spectral type taken
from Tables 67--69 of the Strai\v{z}ys (1992) monograph.  Color excesses
were transformed to extinctions by the equation $A_V = 4.16\,E_{Y-V}$.
Distances $d$ to the stars in parsecs were calculated by the equation
$5\log\,d = V - M_V + 5 - A_V$.  Here $V$ are from Paper I and $M_V$ are
from the tabulation given in Strai\v{z}ys (1992, Appendix I), adjusted
to a Hyades distance modulus of 3.3 mag.  The results are given in Table
1 which gives the star number in the catalog of Paper I, $V$ magnitude,
spectral type and its quality, absolute magnitude $M_V$, interstellar
extinction $A_V$, distance $d$ and the name of the subarea to which the
star is attributed.  The subareas are described lower in this section
and are shown in Figure 2.

\noindent
{\small\tabcolsep=9pt
\begin{longtable}{rrlcccrc}
\multicolumn{8}{c}{\parbox{110mm}{\baselineskip=9pt
{\normbf\ \ \ Table 1.}{\norm\ Stars in the investigated area with most reliable spectral
types determined from {\it Vilnius} photometry. The column $p$ gives the
accuracy estimates of spectral types.\lstrut}}}\\
\hline
\noalign{\vskip1.5mm}
  No. &  $V$~ &  Sp &  $p$ & $M_V$ & $A_V$ & $d$\,(pc) &
\multicolumn{1}{c}{Subarea} \\
\noalign{\vskip1.5mm}
\hline
\noalign{\vskip1mm}
\endfirsthead
\multicolumn{8}{l}{{\normbf\ \ Table 1.}{\norm\ Continued\lstrut}}\\
\hline
\noalign{\vskip1.5mm}
  No. &  $V$~ &  Sp &  $p$ & $M_V$ & $A_V$ & $d$\,(pc) &
\multicolumn{1}{c}{Subarea} \\
\noalign{\vskip1.5mm}
\hline
\noalign{\vskip1mm}
\endhead
\endfoot
   5 &  15.75 & g2.5 V      &  1.0 &  4.59 &  1.36 &   910 &  I  \\
   6 &  12.91 & g8 III      &  1.0 &  0.78 &  1.25 &  1500 &  I  \\
  11 &  15.77 & f6 V        &  0.9 &  3.73 &  1.13 &  1520 &  I  \\
  12 &  15.39 & f5 III      &  0.9 &  1.79 &  1.07 &  3200 &  I  \\
  14 &  15.70 & k0.5 V      &  0.8 &  5.95 &  1.31 &   488 &  I  \\
  15 &  14.56 & g0 IV       &  0.9 &  3.10 &  1.09 &  1190 &  I  \\
  16 &  15.29 & g1 V        &  0.8 &  4.76 &  1.14 &   760 &  I  \\
  23 &  15.70 & g7 V        &  1.0 &  5.40 &  1.00 &   720 &  I  \\
  33 &  13.70 & f7 IV       &  0.9 &  2.10 &  1.11 &  1250 &  I  \\
  34 &  11.75 & f8 V        &  0.9 &  3.84 &  0.24 &   342 &  I  \\
  40 &  15.62 & g5.5 V      &  1.0 &  4.91 &  1.73 &   630 &  I  \\
  47 &  15.84 & g2.5 V      &  0.8 &  4.85 &  1.18 &   920 &  I  \\
  50 &  12.56 & g2 V        &  1.0 &  4.47 &  0.79 &   289 &  I  \\
  51 &  14.98 & f8 IV       &  0.9 &  2.40 &  1.13 &  1950 &  I  \\
  53 &  13.65 & f1 III      &  0.9 &  1.20 &  1.20 &  1780 &  I  \\
  54 &  15.41 & g9 III      &  0.8 &  1.35 &  1.18 &  3760 &  I  \\
  56 &  15.68 & g8 III      &  0.8 &  0.61 &  2.03 &  4060 &     \\
  57 &  15.90 & g7 V        &  1.0 &  5.24 &  1.12 &   810 &  I  \\
  63 &  14.33 & k0 III      &  1.0 &  0.50 &  1.49 &  2950 &  IV  \\
  68 &  13.91 & g8 V        &  1.0 &  5.35 &  0.84 &   350 &  I  \\
  71 &  15.26 & f8 IV       &  0.9 &  2.22 &  1.14 &  2390 &  I  \\
  72 &  16.62 & f7 V        &  0.7 &  4.07 &  1.16 &  1900 &  I  \\
  75 &   9.71 & f5 IV       &  0.9 &  2.71 &  0.09 &   241 &  IV  \\
  76 &  12.40 & f3 V        &  0.9 &  3.16 &  0.74 &   500 &  I  \\
  79 &  11.69 & g1 V        &  0.8 &  4.59 &  0.00 &   263 &  I  \\
  82 &  15.19 & g4 V        &  1.0 &  4.71 &  1.00 &   790 &  I  \\
  83 &  14.49 & g1.5 V      &  1.0 &  4.48 &  0.41 &   830 &  I  \\
  84 &  13.90 & f8 IV       &  0.9 &  1.96 &  1.51 &  1220 &  IV  \\
  86 &  15.59 & f5 V        &  0.9 &  3.56 &  0.99 &  1610 &  I  \\
  87 &  13.38 & g9.5 III    &  1.0 &  0.75 &  1.18 &  1950 &  I  \\
  88 &  14.39 & f4 V        &  0.9 &  3.40 &  1.09 &   960 &  I  \\
  89 &  10.15 & f3 V        &  0.9 &  3.24 &  0.01 &   240 &  I  \\
 101 &  13.82 & g8 III      &  1.0 &  1.59 &  1.54 &  1370 &  IV  \\
 103 &  12.42 & k2 III      &  1.0 &  0.70 &  1.05 &  1360 &  I  \\
 104 &  12.38 & f4 III      &  0.9 &  1.96 &  1.31 &   660 &  IV  \\
 105 &  13.13 & k1 IV       &  1.0 &  3.11 &  1.15 &   590 &  I  \\
 108 &  15.80 & g3 V        &  0.8 &  4.87 &  1.06 &   940 &  I  \\
 110 &  15.71 & g1.5 V      &  0.8 &  4.45 &  1.22 &  1020 &  I  \\
 117 &  15.19 & g9.5 III    &  0.8 &  0.70 &  1.85 &  3380 &  IV  \\
 125 &  16.08 & k1.5 V      &  0.7 &  6.37 &  0.86 &   590 &  I  \\
 126 &  14.99 & f5 V        &  0.9 &  3.64 &  1.11 &  1120 &  I  \\
 134 &  14.69 & f5 IV       &  0.9 &  2.23 &  1.66 &  1450 &  IV  \\
 135 &  14.59 & g9 IV       &  1.0 &  2.64 &  1.48 &  1240 &  IV  \\
 136 &  14.63 & k1.2 V      &  1.0 &  5.46 &  0.92 &   447 &  I  \\
 138 &  16.42 & g3 IV       &  0.7 &  2.89 &  1.16 &  2980 &  I  \\
 141 &  16.15 & g1.5 IV     &  0.7 &  3.09 &  1.50 &  2040 &  I  \\
 143 &  15.41 & k1 IV       &  0.8 &  3.64 &  1.38 &  1190 &  I  \\
 144 &  16.18 & g3 IV       &  0.7 &  3.44 &  1.60 &  1690 &  IV  \\
 147 &  15.52 & g8 V        &  1.0 &  5.01 &  1.19 &   730 &  IV  \\
 149 &  16.07 & k0.5 IV     &  0.8 &  3.19 &  1.16 &  2210 &  I  \\
 150 &  14.98 & k0.5 V      &  1.0 &  5.84 &  0.56 &   520 &  I  \\
 153 &  13.03 & k3.5 III    &  1.0 &  0.48 &  1.07 &  1980 &  I  \\
 158 &  15.96 & g9.5 IV     &  0.8 &  2.75 &  1.26 &  2450 &  I  \\
 160 &  14.45 & a5 IV       &  0.9 &  1.24 &  1.78 &  1930 &  IV  \\
 161 &  11.60 & g8 III      &  1.0 &  0.83 &  1.50 &   710 &  I  \\
 162 &  15.04 & k0.5 V      &  1.0 &  5.63 &  0.79 &   530 &  I  \\
 168 &  13.30 & g9.5 III    &  1.0 &  0.54 &  2.13 &  1340 &  IV  \\
 174 &  15.54 & f6 IV       &  0.9 &  2.62 &  1.63 &  1820 &  I  \\
 176 &  16.14 & g1 V        &  0.8 &  4.56 &  0.83 &  1410 &  I  \\
 177 &  15.61 & g3 V        &  0.8 &  4.80 &  1.50 &   730 &  I  \\
 178 &  14.98 & k0.5 IV     &  0.8 &  2.77 &  1.10 &  1660 &  I  \\
 179 &  13.79 & f6 IV       &  0.9 &  2.85 &  1.27 &   860 &  IV  \\
 181 &  15.93 & g2 V        &  0.8 &  4.70 &  1.29 &   970 &  I  \\
 182 &  13.11 & g9 III      &  1.0 &  0.73 &  1.65 &  1400 &  IV  \\
 184 &  15.09 & g1.5 IV     &  1.0 &  3.23 &  1.21 &  1340 &  I  \\
 185 &  15.32 & k0 III      &  0.8 &  1.08 &  1.58 &  3400 &  IV  \\
 188 &  15.09 & k3.2 III    &  0.8 &  0.73 &  1.25 &  4170 &     \\
 190 &   9.80 & k3 III      &  1.0 &  0.29 &  1.62 &   720 &  I  \\
 194 &  11.56 & g3 V        &  1.0 &  4.82 &  0.15 &   209 &  IV  \\
 199 &  15.45 & g1 IV       &  0.7 &  3.02 &  1.85 &  1310 &  IV  \\
 200 &  12.78 & k0.7 III    &  1.0 &  0.79 &  1.67 &  1150 &  IV  \\
 202 &  15.58 & g7 V        &  1.0 &  5.47 &  1.23 &   600 &  I  \\
 204 &  15.22 & g0 V        &  1.0 &  4.00 &  1.45 &   900 &  IV  \\
 205 &  14.52 & g3 V        &  1.0 &  4.87 &  1.08 &   520 &  I  \\
 207 &  13.08 & g0 IV       &  0.9 &  2.54 &  0.97 &   820 &  IV  \\
 208 &  15.67 & k1 IV       &  0.8 &  3.66 &  1.34 &  1370 &  IV  \\
 209 &  14.36 & g3 V        &  1.0 &  4.86 &  0.86 &   530 &  I  \\
 215 &  12.61 & f3 IV       &  0.9 &  2.54 &  0.83 &   710 &  I  \\
 216 &  15.68 & g1.5 V      &  1.0 &  4.44 &  1.34 &   960 &  I  \\
 217 &  14.56 & f8 IV       &  0.9 &  2.38 &  1.53 &  1350 &  IV  \\
 223 &  15.53 & g5 IV       &  0.7 &  3.04 &  1.36 &  1680 &  I  \\
 225 &  14.50 & g9 IV       &  1.0 &  3.48 &  1.15 &   940 &  I  \\
 226 &  16.47 & f5 V        &  0.7 &  3.66 &  1.41 &  1900 &  IV  \\
 228 &  13.64 & g8 IV       &  1.0 &  3.01 &  1.47 &   680 &  IV  \\
 231 &  15.02 & g2 IV       &  1.0 &  3.02 &  1.63 &  1190 &  IV  \\
 233 &  15.41 & k2.5 V      &  0.8 &  5.92 &  0.49 &   630 &  I  \\
 235 &  11.72 & k0 III      &  1.0 &  0.72 &  1.16 &   930 &  IV  \\
 241 &  15.47 & g6 III      &  0.7 &  1.01 &  2.15 &  2890 &  IV  \\
 242 &  13.66 & g0 V        &  1.0 &  4.20 &  0.65 &   580 &  I  \\
 249 &  14.67 & k0.5 III    &  0.8 &  0.59 &  1.93 &  2700 &  IV  \\
 250 &  14.49 & k1.7 III    &  0.8 &  0.67 &  1.56 &  2820 &  IV  \\
 251 &  13.49 & g2.5 V      &  1.0 &  4.40 &  1.09 &   400 &  I  \\
 252 &  15.45 & g9.5 V      &  1.0 &  5.36 &  1.20 &   600 &  IV  \\
 254 &  14.00 & k3 V        &  0.8 &  6.63 &  0.71 &   214 &  IV  \\
 258 &  14.16 & f7 IV       &  0.9 &  2.70 &  1.39 &  1030 &  IV  \\
 259 &  13.52 & f7 V        &  0.8 &  3.71 &  0.67 &   670 &  I  \\
 266 &   7.70 & k3 III      &  1.0 &  0.70 &  0.28 &   221 &  IV  \\
 267 &  14.98 & g9 IV       &  1.0 &  3.51 &  1.43 &  1020 &  IV  \\
 274 &  15.78 & g2 V        &  1.0 &  4.59 &  1.45 &   890 &  I  \\
 275 &  12.98 & g5.5 III    &  1.0 &  0.71 &  1.54 &  1400 &  IV  \\
 277 &  15.06 & g9.5 IV     &  0.8 &  2.90 &  1.70 &  1240 &  I  \\
 284 &  15.52 & g8 V        &  1.0 &  5.20 &  1.36 &   620 &  I  \\
 285 &  12.51 & g8 V        &  1.0 &  5.24 &  0.08 &   274 &  I  \\
 286 &  13.56 & k3.5 III    &  0.8 &  0.45 &  1.75 &  1880 &  I  \\
 290 &  14.45 & k0.7 IV     &  1.0 &  2.63 &  1.94 &   950 &  IV  \\
 292 &  15.35 & g9 V        &  1.0 &  5.67 &  1.43 &   448 &  IV  \\
 293 &  14.97 & f7 IV       &  0.9 &  2.37 &  1.21 &  1890 &  I  \\
 300 &  11.65 & f6 III      &  0.9 &  2.00 &  0.76 &   600 &  I  \\
 301 &  14.19 & g9.5 IV     &  1.0 &  2.48 &  1.32 &  1200 &  IV  \\
 302 &  12.51 & f5 IV       &  0.9 &  1.77 &  1.19 &   820 &  IV  \\
 304 &  11.37 & g6 V        &  1.0 &  5.14 &  0.13 &   165 &  IV  \\
 305 &  14.15 & f6 IV       &  0.9 &  2.52 &  1.62 &  1010 &  IV  \\
 307 &  16.09 & f8 V        &  0.8 &  4.15 &  1.11 &  1460 &  I  \\
 308 &  12.53 & f7 IV       &  0.9 &  2.74 &  1.29 &   500 &  I  \\
 309 &  13.42 & g5.5 V      &  1.0 &  4.83 &  0.74 &   371 &  I  \\
 310 &  15.51 & k0.5 V      &  1.0 &  5.92 &  1.31 &   453 &  IV  \\
 312 &  14.74 & k2 III      &  0.8 &  1.16 &  2.15 &  1930 &  IV  \\
 314 &  15.68 & g3 V        &  0.8 &  4.71 &  1.46 &   800 &  IV  \\
 315 &  15.65 & k0 V        &  0.8 &  5.77 &  1.32 &   520 &  IV  \\
 321 &  13.06 & f0 IV       &  0.9 &  2.11 &  1.23 &   880 &  IV  \\
 323 &  14.12 & f9.5 IV     &  0.9 &  2.94 &  1.36 &   920 &  IV  \\
 324 &  14.70 & g1.5 IV     &  1.0 &  2.51 &  1.10 &  1660 &  IV  \\
 325 &  15.94 & f7 V        &  0.7 &  3.94 &  2.09 &   960 &  IV  \\
 331 &  14.07 & g9.5 IV     &  1.0 &  2.89 &  2.46 &   560 &  IV  \\
 333 &  15.26 & f9 IV       &  0.9 &  2.36 &  1.44 &  1960 &  IV  \\
 335 &  15.81 & f9.5 V      &  0.7 &  4.36 &  1.65 &   910 &  IV  \\
 337 &  15.35 & f9.5 V      &  0.9 &  4.23 &  1.11 &  1000 &  I  \\
 339 &  13.78 & k0 III      &  1.0 &  0.95 &  1.53 &  1820 &  IV  \\
 348 &  13.93 & a8 III      &  0.9 &  1.02 &  1.29 &  2110 &  IV  \\
 349 &  14.48 & g9.5 III    &  1.0 &  1.00 &  1.62 &  2350 &  I  \\
 350 &  15.55 & f7 IV       &  0.9 &  2.68 &  2.02 &  1480 &  IV  \\
 351 &  11.44 & g9 III      &  1.0 &  0.27 &  1.61 &   820 &  IV  \\
 352 &  15.35 & f9 IV       &  0.9 &  2.71 &  1.92 &  1400 &  IV  \\
 353 &  15.25 & g2.5 V      &  1.0 &  4.60 &  1.63 &   640 &  IV  \\
 356 &  12.41 & k0.5 III    &  1.0 &  0.73 &  1.55 &  1060 &  IV  \\
 360 &  14.05 & b7 V        &  1.0 & -0.11 &  2.20 &  2460 &  IV  \\
 362 &  15.77 & g0 V        &  0.8 &  4.45 &  1.57 &   890 &  IV  \\
 363 &  14.39 & g9.5 IV     &  1.0 &  2.77 &  2.05 &   820 &  IV  \\
 369 &  12.68 & f4 IV       &  0.9 &  2.38 &  0.97 &   740 &  IV  \\
 373 &  14.02 & g2.5 IV     &  1.0 &  3.01 &  1.68 &   740 &  IV  \\
 377 &  13.10 & a7 IV       &  0.9 &  1.34 &  1.82 &   970 &  IV  \\
 378 &  15.20 & k0.7 III    &  0.8 &  0.78 &  1.57 &  3710 &  IV  \\
 380 &  14.49 & f7 V        &  0.9 &  3.95 &  0.85 &   870 &  I  \\
 383 &  14.87 & k1.2 III    &  0.8 &  0.33 &  2.55 &  2500 &  IV  \\
 384 &  14.94 & g1 V        &  0.9 &  4.25 &  1.23 &   780 &  IV  \\
 386 &  15.57 & g4 V        &  1.0 &  4.77 &  1.24 &   820 &     \\
 391 &  14.67 & k7 V        &  0.8 &  8.15 &  0.00 &   201 &  I  \\
 394 &  15.21 & g1 IV       &  1.0 &  2.81 &  1.51 &  1500 &  IV  \\
 395 &  12.55 & g6 V        &  1.0 &  5.14 &  0.11 &   288 &  IV  \\
 398 &  15.45 & f9 V        &  0.9 &  4.13 &  1.39 &   970 &  IV  \\
 400 &  13.51 & f8 IV       &  0.9 &  2.68 &  1.01 &   920 &  IV  \\
 403 &  12.80 & f3 III      &  0.8 &  1.60 &  1.46 &   890 &  IV  \\
 409 &  14.75 & k5 V        &  1.0 &  7.16 &  0.45 &   268 &  IV  \\
 411 &  13.37 & f8 IV       &  0.9 &  1.88 &  1.18 &  1150 &  IV  \\
 415 &  15.65 & g1 IV       &  0.9 &  2.21 &  1.39 &  2560 &  IV  \\
 418 &  12.75 & f1 IV       &  0.9 &  1.89 &  1.24 &   840 &  IV  \\
 422 &  15.23 & g3 V        &  1.0 &  4.86 &  1.27 &   660 &  IV  \\
 423 &  15.65 & g1 IV       &  0.5 &  3.04 &  1.76 &  1480 &  IV  \\
 424 &  14.68 & f9 IV       &  0.9 &  1.49 &  1.51 &  2170 &  IV  \\
 426 &  11.18 & f4 III      &  0.9 &  1.82 &  1.37 &   395 &  V  \\
 428 &  12.22 & a5 IV       &  0.9 &  0.93 &  1.33 &   990 &  IV  \\
 431 &  13.33 & k2.2 V      &  0.8 &  6.25 &  0.15 &   243 &  V  \\
 434 &  15.36 & g2.5 IV     &  0.9 &  2.53 &  1.56 &  1800 &  IV  \\
 437 &  16.43 & f7 V        &  0.7 &  3.91 &  1.60 &  1520 &  IV  \\
 439 &  10.70 & f0 IV       &  0.9 &  1.84 &  1.04 &   366 &  IV  \\
 443 &  14.83 & f7 IV       &  0.9 &  2.43 &  1.42 &  1570 &  IV  \\
 446 &  15.57 & g6 IV       &  0.8 &  3.17 &  1.45 &  1550 &  IV  \\
 448 &  11.71 & g9 III      &  1.0 &  0.73 &  1.56 &   770 &  IV  \\
 451 &  14.00 & f7 IV       &  0.9 &  2.82 &  1.24 &   970 &  IV  \\
 456 &  15.60 & k0 III      &  0.8 &  0.68 &  1.92 &  3990 &  IV  \\
 457 &  15.35 & g0 IV       &  0.9 &  2.75 &  1.42 &  1720 &  IV  \\
 460 &  16.35 & g2 V        &  0.7 &  4.44 &  1.50 &  1210 &  IV  \\
 464 &  14.96 & m2.5 V      &  0.8 & 10.06 &  0.25 &    85 &  IV  \\
 466 &  16.55 & g7 V        &  0.8 &  5.40 &  1.24 &   960 &  IV  \\
 469 &  14.49 & k0 IV       &  1.0 &  3.56 &  1.27 &   860 &  IV  \\
 473 &  12.23 & g0 V        &  0.9 &  4.33 &  0.18 &   349 &  V  \\
 476 &  13.74 & g7 V        &  1.0 &  5.37 &  0.64 &   352 &  IV  \\
 477 &  12.57 & f5 V        &  0.9 &  2.17 &  1.29 &   660 &  IV  \\
 480 &  13.74 & k3.5 III    &  0.7 &  0.65 &  1.53 &  2060 &      \\
 482 &  15.66 & g6 III      &  0.8 &  0.85 &  2.10 &  3480 &  IV  \\
 485 &  13.09 & k0.7 IV     &  1.0 &  3.12 &  0.94 &   640 &  I  \\
 494 &  16.24 & g2.5 V      &  0.7 &  4.71 &  1.93 &   830 &  IV  \\
 495 &  15.22 & g1 V        &  0.9 &  4.51 &  0.74 &   990 &  I  \\
 496 &  13.92 & g1 IV       &  1.0 &  2.27 &  1.56 &  1040 &  IV  \\
 497 &  15.82 & m2 V        &  0.8 &  7.37 &  1.83 &   211 &  IV  \\
 498 &  11.02 & f4 V        &  0.9 &  3.45 &  0.14 &   307 &  V  \\
 499 &  12.81 & m0 III      &  1.0 & -0.68 &  1.63 &  2360 &  IV  \\
 500 &  13.65 & f9 IV       &  0.9 &  2.71 &  1.08 &   940 &  I  \\
 501 &  11.95 & f3 IV       &  0.9 &  2.37 &  0.85 &   560 &  I  \\
 504 &  15.65 & f8 IV       &  0.8 &  2.72 &  1.78 &  1700 &  V  \\
 506 &  15.15 & k0.5 IV     &  0.8 &  3.26 &  1.39 &  1260 &  IV  \\
 511 &  16.31 & f7 V        &  0.7 &  3.90 &  1.44 &  1560 &  I  \\
 513 &  16.00 & f9.5 V      &  0.7 &  4.37 &  1.66 &   980 &  IV  \\
 514 &  15.98 & g8 V        &  1.0 &  5.56 &  0.91 &   800 &  I  \\
 516 &  15.82 & f6 V        &  0.7 &  3.94 &  1.98 &   960 &  IV  \\
 519 &  15.33 & g5.5 V      &  1.0 &  3.05 &  1.42 &   670 &  IV  \\
 521 &  14.56 & k2.5 V      &  0.8 &  6.45 &  0.67 &   309 &  I  \\
 523 &  12.54 & k1.5 III    &  1.0 &  0.90 &  1.34 &  1150 &  IV  \\
 527 &  14.30 & f6 V        &  0.9 &  3.26 &  1.14 &   950 &  I  \\
 531 &  15.71 & f8 V        &  0.5 &  4.21 &  1.96 &   810 &  V  \\
 533 &  13.61 & g0 IV       &  1.0 &  2.38 &  1.44 &   910 &  IV  \\
 537 &  16.18 & k3 V        &  0.8 &  6.56 &  0.99 &   530 &  IV  \\
 538 &  15.91 & g3 V        &  0.7 &  4.51 &  2.36 &   640 &  V  \\
 540 &  15.16 & g9.5 V      &  1.0 &  5.55 &  0.86 &   560 &  I  \\
 541 &  13.20 & k0.7 V      &  1.0 &  5.77 &  0.27 &   272 &  IV  \\
 544 &  12.27 & g2 V        &  0.8 &  4.69 &  0.12 &   311 &  IV  \\
 545 &  15.50 & f8 V        &  0.7 &  4.11 &  2.11 &   720 &  IV  \\
 554 &  12.86 & k3 III      &  1.0 &  0.67 &  1.22 &  1560 &  I  \\
 560 &  14.89 & f7 V        &  0.8 &  3.92 &  1.25 &   880 &  IV  \\
 562 &  14.30 & g2.5 III    &  1.0 &  0.84 &  1.95 &  2000 &  V  \\
 563 &  14.04 & k1 V        &  1.0 &  6.03 &  0.44 &   327 &  V  \\
 565 &  14.97 & g5 V        &  1.0 &  4.81 &  1.25 &   610 &  IV  \\
 567 &  15.73 & a9 IV       &  0.8 &  1.97 &  1.86 &  2400 &  V  \\
 569 &  13.99 & k2.2 III    &  0.8 &  0.50 &  2.71 &  1430 &  V  \\
 570 &  14.10 & k0 IV       &  1.0 &  2.82 &  1.13 &  1070 &  I  \\
 572 &  14.56 & f6 V        &  0.7 &  3.79 &  1.00 &   900 &  I  \\
 574 &  12.44 & k2.5 III    &  0.7 &  1.41 &  1.11 &   970 &  I  \\
 576 &  15.01 & g8 III      &  0.8 &  0.89 &  3.07 &  1630 &  V  \\
 577 &  14.04 & k0 V        &  0.8 &  6.02 &  0.50 &   320 &  V  \\
 578 &  16.23 & g1.5 V      &  0.7 &  4.46 &  1.30 &  1240 &  I  \\
 581 &  15.48 & f8 IV       &  0.9 &  2.37 &  1.49 &  2100 &  I  \\
 582 &  15.43 & g7 III      &  0.8 &  0.71 &  1.57 &  4280 &  I  \\
 583 &  14.29 & g1.5 V      &  0.8 &  4.64 &  0.80 &   590 &  I  \\
 595 &  13.18 & k2 V        &  1.0 &  6.34 &  1.56 &   114 &  V  \\
 597 &  13.31 & k0.5 IV     &  1.0 &  2.54 &  1.46 &   730 &  I  \\
 598 &  15.53 & k2 V        &  0.8 &  6.26 &  0.85 &   483 &  I  \\
 602 &  16.56 & f7 IV       &  0.7 &  2.21 &  1.86 &  3150 &  I  \\
 604 &  15.67 & f9 IV       &  0.9 &  2.95 &  1.53 &  1730 &  IV  \\
 607 &  14.73 & f7 IV       &  0.9 &  1.88 &  1.98 &  1490 &  IV  \\
 609 &  14.57 & k1.2 V      &  1.0 &  5.47 &  1.63 &   314 &  V  \\
 611 &   9.11 & m0 III      &  1.0 & -0.66 &  1.10 &   540 &  I  \\
 615 &  14.96 & a1 IV       &  1.0 &  0.47 &  1.51 &  3940 &  II  \\
 622 &  13.47 & k0 III      &  0.7 &  0.38 &  2.57 &  1270 &  V  \\
 623 &  14.17 & k1 IV       &  0.8 &  2.54 &  1.75 &   950 &  II  \\
 626 &  12.83 & f5 IV       &  0.9 &  2.36 &  1.73 &   560 &  V  \\
 632 &  11.82 & k0.7 V      &  1.0 &  6.00 &  0.20 &   133 &  V  \\
 633 &  15.36 & g8.5 V      &  1.0 &  5.66 &  1.50 &   437 &  V  \\
 634 &  13.95 & g8.5 IV     &  1.0 &  3.16 &  1.44 &   740 &  II  \\
 636 &  14.57 & k6 V        &  0.6 &  7.83 &  0.55 &   173 &  V  \\
 638 &  13.05 & k1 V        &  1.0 &  6.12 &  0.18 &   224 &  II  \\
 640 &   9.76 & k2.2 III    &  0.8 &  0.82 &  0.83 &   420 &  II  \\
 643 &  12.94 & g4 V        &  1.0 &  4.94 &  0.30 &   346 &  II  \\
 644 &  16.15 & g1 IV       &  0.7 &  3.76 &  1.68 &  1390 &  IV  \\
 649 &  12.10 & k3.7 III    &  1.0 &  0.27 &  1.70 &  1070 &  II  \\
 652 &  13.82 & k3.2 V      &  0.8 &  6.69 &  0.18 &   245 &  V  \\
 656 &  13.67 & g4 V        &  1.0 &  4.55 &  0.70 &   485 &  II  \\
 657 &  12.95 & k0 V        &  1.0 &  5.57 &  0.15 &   279 &  V  \\
 659 &  14.53 & g9.5 III    &  0.8 &  1.23 &  1.98 &  1840 &  IV  \\
 660 &  11.67 & g4 V        &  1.0 &  4.95 &  0.19 &   202 &  V  \\
 662 &  16.55 & f8 V        &  0.7 &  3.84 &  1.78 &  1530 &  IV  \\
 669 &  14.23 & f7 V        &  0.9 &  3.74 &  0.95 &   810 &  II  \\
 671 &  14.24 & k1.2 III    &  0.8 &  1.07 &  2.20 &  1560 &  IV  \\
 674 &  15.54 & g5 IV       &  1.0 &  3.48 &  1.22 &  1470 &  IV  \\
 676 &  16.05 & g8 V        &  0.8 &  5.43 &  1.28 &   740 &  IV  \\
 682 &  14.06 & g8 V        &  1.0 &  5.13 &  0.58 &   469 &  II  \\
 693 &  15.01 & f3 V        &  0.6 &  3.14 &  2.53 &   740 &  V  \\
 694 &  14.28 & f6 III      &  0.9 &  1.58 &  1.81 &  1510 &  IV  \\
 695 &  14.93 & f6 IV       &  0.9 &  2.16 &  1.66 &  1660 &  IV  \\
 696 &  15.34 & g8 V        &  1.0 &  5.06 &  0.84 &   770 &  IV  \\
 698 &  14.34 & g9 III      &  1.0 &  0.71 &  1.89 &  2230 &  IV  \\
 700 &  12.58 & f4 IV       &  0.9 &  2.56 &  0.73 &   720 &  II  \\
 703 &  13.86 & f4 IV       &  0.9 &  2.43 &  1.09 &  1160 &  II  \\
 705 &  15.04 & f7 IV       &  0.9 &  2.54 &  1.33 &  1710 &  II  \\
 706 &  15.75 & f4 V        &  0.8 &  3.31 &  1.19 &  1780 &  II  \\
 708 &  11.81 & f9 V        &  0.8 &  4.15 &  0.18 &   312 &  V  \\
 709 &  15.08 & g6 V        &  1.0 &  4.92 &  0.82 &   740 &  II  \\
 710 &  15.85 & f9.5 V      &  0.9 &  4.26 &  1.39 &  1100 &  IV  \\
 711 &  14.76 & f8 IV       &  0.9 &  1.88 &  2.11 &  1420 &  IV  \\
 712 &  14.60 & f8 IV       &  0.9 &  2.86 &  1.15 &  1310 &  IV  \\
 714 &  12.18 & k2.7 V      &  1.0 &  6.47 &  0.29 &   121 &  V  \\
 719 &  15.21 & f7 IV       &  0.9 &  2.00 &  1.29 &  2420 &  II  \\
 720 &  16.10 & g1 V        &  0.7 &  4.45 &  1.76 &   950 &  IV \\
 722 &  15.21 & g2 V        &  1.0 &  4.67 &  0.44 &  1050 &  II  \\
 727 &  14.93 & k0.5 V      &  1.0 &  5.89 &  0.73 &   460 &  II  \\
 731 &  12.66 & k2 V        &  1.0 &  6.33 &  0.23 &   166 &  IV  \\
 738 &  15.76 & g8 IV       &  0.8 &  2.65 &  1.90 &  1740 &  IV  \\
 739 &  16.01 & k1.2 V      &  0.8 &  6.22 &  0.90 &   600 &  IV  \\
 742 &  15.61 & g3 V        &  0.8 &  4.75 &  1.24 &   840 &  IV  \\
 750 &  14.49 & f6 IV       &  0.9 &  2.69 &  1.54 &  1130 &  IV  \\
 754 &  11.69 & k4.2 III    &  1.0 &  0.21 &  1.16 &  1160 &  II  \\
 757 &  12.77 & a5 IV       &  0.9 &  1.32 &  1.42 &  1010 &  IV  \\
 760 &  14.01 & f6 IV       &  0.9 &  2.42 &  1.15 &  1220 &  IV  \\
 769 &  15.88 & g4 V        &  1.0 &  5.02 &  1.34 &   800 &  IV  \\
 770 &  12.95 & f8 IV       &  0.9 &  2.75 &  0.88 &   730 &  IV  \\
 775 &  16.36 & g2 V        &  0.7 &  4.80 &  1.39 &  1080 &  IV  \\
 776 &  15.88 & f6 III      &  0.7 &  1.54 &  2.13 &  2770 &  IV  \\
 778 &  13.81 & f7 IV       &  0.9 &  2.73 &  0.95 &  1060 &  IV  \\
 780 &  14.27 & k0 III      &  1.0 &  0.81 &  2.01 &  1940 &  IV  \\
 786 &  16.28 & g0 IV       &  0.7 &  3.10 &  1.74 &  1940 &  IV  \\
 789 &  13.86 & f7 IV       &  0.9 &  2.05 &  1.07 &  1400 &  IV  \\
 793 &  15.87 & f9 V        &  0.8 &  4.04 &  1.39 &  1230 &  IV  \\
 797 &  14.05 & f6 IV       &  0.9 &  2.60 &  1.39 &  1030 &  IV  \\
 798 &  14.35 & k0.7 V      &  1.0 &  5.73 &  0.36 &   447 &  II  \\
 799 &  15.09 & g5.5 III    &  1.0 &  0.80 &  1.58 &  3490 &  II  \\
 800 &  15.79 & f4 III      &  0.7 &  1.92 &  2.02 &  2350 &  IV  \\
 803 &  14.71 & f0 IV       &  0.9 &  2.02 &  1.51 &  1720 &  IV  \\
 804 &  14.04 & g4 V        &  1.0 &  4.15 &  0.17 &   880 &  II  \\
 805 &  14.72 & g9.5 V      &  1.0 &  5.63 &  0.84 &   447 &  IV  \\
 810 &  14.57 & a8 IV       &  0.9 &  1.70 &  2.01 &  1490 &  IV  \\
 811 &  12.13 & g2.5 IV     &  1.0 &  3.35 &  0.51 &   449 &  IV  \\
 812 &  15.53 & g2.5 V      &  1.0 &  4.61 &  1.21 &   870 &  IV  \\
 814 &  15.96 & f7 IV       &  0.7 &  2.48 &  1.93 &  2050 &  IV  \\
 815 &  16.01 & g2 V        &  0.7 &  4.78 &  1.42 &   920 &  IV  \\
 820 &  14.63 & g1 V        &  0.9 &  4.45 &  0.85 &   730 &  IV  \\
 821 &   8.99 & f8 V        &  0.8 &  4.03 &  0.08 &    95 &  IV  \\
 823 &  15.93 & k2.5 V      &  0.8 &  6.35 &  1.00 &   520 &  IV  \\
 825 &  13.74 & g1.5 V      &  0.9 &  4.56 &  0.66 &   510 &  III  \\
 826 &  15.44 & k1.2 V      &  1.0 &  5.72 &  1.00 &   560 &  IV  \\
 828 &  14.09 & k0 IV       &  1.0 &  2.64 &  1.82 &   840 &  IV  \\
 831 &  11.42 & f8 V        &  0.8 &  4.03 &  0.15 &   281 &  IV  \\
 835 &  14.94 & f9 IV       &  0.9 &  2.10 &  1.42 &  1920 &  IV  \\
 839 &  14.33 & f8 V        &  0.9 &  3.96 &  0.61 &   900 &  IV \\
 844 &  13.03 & g8 III      &  1.0 &  0.84 &  1.15 &  1610 &  II  \\
 846 &  14.70 & f9 IV       &  0.9 &  2.10 &  1.54 &  1630 &  III  \\
 847 &  15.61 & g1 V        &  1.0 &  4.41 &  0.94 &  1130 &  IV  \\
 855 &  15.71 & g0 IV       &  0.9 &  2.57 &  1.51 &  2120 &  IV  \\
 857 &  13.53 & f9 V        &  0.9 &  4.05 &  0.58 &   600 &  III  \\
 861 &  14.91 & g0 IV       &  1.0 &  2.85 &  1.83 &  1110 &  II  \\
 868 &  12.05 & g8.5 III    &  1.0 &  0.75 &  1.18 &  1060 &  IV  \\
 871 &  12.02 & f7 IV       &  0.9 &  3.00 &  0.79 &   442 &  IV  \\
 876 &  15.34 & f7 IV       &  0.8 &  2.27 &  1.73 &  1860 &  IV  \\
 881 &  14.76 & f5 IV       &  0.9 &  2.63 &  1.37 &  1420 &  IV  \\
 887 &  11.51 & g1 V        &  0.9 &  4.35 &  0.02 &   268 &  III  \\
 893 &  15.48 & f8 V        &  0.8 &  3.91 &  1.09 &  1250 &  II  \\
 895 &  14.99 & f4 V        &  0.9 &  3.49 &  0.96 &  1280 &  IV  \\
 896 &  14.10 & k0.5 III    &  1.0 &  0.78 &  1.63 &  2190 &  IV  \\
 903 &  12.54 & f8 IV       &  0.9 &  2.32 &  1.10 &   670 &      \\
 905 &  15.24 & f9 V        &  0.9 &  4.08 &  0.67 &  1250 &  II  \\
 906 &  14.52 & g4 V        &  1.0 &  5.07 &  0.79 &   540 &  II  \\
 910 &  13.79 & f3 III      &  0.8 &  0.90 &  1.60 &  1810 &  III  \\
 911 &  13.69 & f4 IV       &  0.9 &  2.44 &  1.12 &  1060 &  II  \\
 915 &  13.83 & g1 IV       &  1.0 &  2.77 &  0.94 &  1060 &  II  \\
 917 &  15.52 & g6 V        &  1.0 &  5.14 &  0.55 &   920 &  IV  \\
 923 &  14.53 & g0 V        &  1.0 &  4.55 &  0.66 &   730 &  II  \\
 924 &  15.36 & f7 IV       &  0.9 &  2.61 &  1.19 &  2050 &  II  \\
 925 &  14.64 & a8 IV       &  1.0 &  1.66 &  1.76 &  1760 &      \\
 926 &  15.93 & f8 V        &  0.8 &  3.97 &  0.87 &  1660 &  II  \\
 928 &  13.04 & k1.2 III    &  1.0 &  0.80 &  1.07 &  1710 &  II  \\
 932 &  14.03 & g6 V        &  1.0 &  4.78 &  0.46 &   570 &  II  \\
 935 &  14.69 & g4 V        &  1.0 &  4.74 &  0.57 &   750 &  III \\
 936 &  14.21 & g5.5 IV     &  1.0 &  2.41 &  1.38 &  1210 &  II  \\
 939 &  14.40 & k0.5 V      &  1.0 &  5.84 &  0.59 &   392 &  II  \\
 943 &  13.06 & g7 III      &  1.0 &  0.77 &  1.78 &  1260 &  III  \\
 946 &  13.48 & g2 V        &  1.0 &  4.42 &  0.37 &   550 &  III  \\
 950 &  14.92 & g3 V        &  1.0 &  4.86 &  0.57 &   790 &  II  \\
 952 &  11.49 & f7 IV       &  0.9 &  2.52 &  0.37 &   530 &  III  \\
 953 &  11.40 & f5 V        &  0.9 &  3.56 &  0.28 &   324 &  II  \\
 954 &  15.35 & k2.7 V      &  1.0 &  6.58 &  0.77 &   398 &  III  \\
 956 &  10.90 & g0 V        &  0.8 &  4.18 &  0.15 &   206 &  III  \\
 957 &  14.28 & g0 V        &  0.8 &  4.35 &  0.68 &   710 &  II  \\
 958 &  15.95 & g2.5 V      &  0.8 &  4.69 &  1.31 &   970 &  III  \\
 959 &  13.90 & g3 IV       &  0.8 &  3.36 &  0.54 &  1000 &  II  \\
 962 &  16.10 & f8 IV       &  0.7 &  1.85 &  2.11 &  2670 &  IV  \\
 963 &  13.38 & k2.5 III    &  1.0 &  1.11 &  1.36 &  1520 &  IV  \\
 966 &  13.59 & a8 IV       &  0.9 &  1.60 &  1.52 &  1240 &      \\
 971 &  14.78 & f5 IV       &  0.9 &  2.57 &  1.23 &  1570 &  II  \\
 972 &  12.51 & g6 V        &  1.0 &  5.15 &  0.15 &   276 &  II  \\
 973 &  12.63 & k0.5 III    &  1.0 &  0.10 &  1.85 &  1370 &  III  \\
 977 &  14.66 & k1.7 V      &  0.8 &  6.09 &  0.35 &   440 &  II  \\
 978 &  15.32 & f5 IV       &  0.9 &  2.68 &  1.38 &  1790 &  II  \\
 980 &  13.76 & g1 V        &  0.9 &  4.41 &  0.54 &   580 &  III  \\
 983 &  14.85 & k0.5 V      &  1.0 &  6.00 &  0.47 &   473 &  III  \\
 984 &  13.81 & f8 IV       &  0.9 &  2.71 &  1.45 &   850 &  III  \\
 987 &  15.45 & g9.5 V      &  1.0 &  5.54 &  0.89 &   640 &  III  \\
 988 &  14.11 & f8 V        &  0.9 &  4.00 &  0.46 &   850 &  II  \\
 989 &  10.28 & a7 V        &  0.9 &  2.29 &  0.60 &   300 &  II  \\
 991 &  15.13 & g1.5 V      &  1.0 &  4.43 &  0.94 &   900 &  II  \\
 992 &  15.86 & k2 V        &  0.7 &  6.40 &  0.78 &   550 &  IV \\
 993 &  15.23 & g8.5 IV     &  1.0 &  3.49 &  1.06 &  1370 &  II  \\
 995 &  14.48 & g7 V        &  1.0 &  5.18 &  0.30 &   630 &  II  \\
 996 &  16.01 & k1.2 V      &  0.8 &  5.51 &  0.35 &  1070 &  II  \\
 998 &  14.42 & g8.5 V      &  1.0 &  5.56 &  0.53 &   463 &  II  \\
1000 &  12.50 & f5 V        &  0.9 &  3.50 &  0.52 &   495 &  III \\
1001 &  12.63 & k0.5 V      &  1.0 &  5.74 &  0.09 &   229 &  II  \\
1007 &  15.04 & g1 IV       &  0.9 &  3.12 &  0.77 &  1700 &  II  \\
1010 &  14.74 & k0 IV       &  1.0 &  2.73 &  1.30 &  1390 &  II  \\
1012 &  11.77 & g4 V        &  1.0 &  4.94 &  0.14 &   218 &      \\
1013 &  15.41 & g9 III      &  0.8 &  1.22 &  1.81 &  2990 &  III  \\
1017 &  14.86 & f9 V        &  0.8 &  4.47 &  0.74 &   850 &  II  \\
1018 &  11.54 & f5 V        &  0.9 &  3.43 &  0.38 &   352 &  III  \\
1020 &  16.67 & f7 V        &  0.7 &  4.09 &  1.23 &  1860 &  II  \\
1021 &  14.45 & g8 III      &  1.0 &  0.88 &  2.06 &  2000 &  III  \\
1023 &  14.21 & g8.5 IV     &  1.0 &  3.17 &  1.16 &   950 &  II  \\
1024 &  12.38 & k5.5 V      &  0.8 &  7.37 &  0.22 &    91 &  III  \\
1026 &  14.90 & g8 IV       &  1.0 &  3.07 &  1.83 &  1000 &  III  \\
1029 &  15.85 & k2.7 V      &  0.8 &  6.52 &  0.59 &   560 &  II  \\
1031 &  13.31 & f8 IV       &  0.9 &  2.57 &  1.03 &   880 &      \\
1032 &  11.83 & f8 IV       &  0.9 &  2.92 &  0.31 &   520 &  III  \\
1033 &  13.68 & k0.5 III    &  1.0 &  0.82 &  1.60 &  1780 &  III  \\
1040 &  16.30 & f9.5 V      &  0.8 &  4.40 &  0.98 &  1530 &  II  \\
1041 &  15.39 & g7 III      &  0.8 &  0.15 &  2.06 &  4320 &  III  \\
1043 &  16.50 & g4 V        &  0.7 &  5.20 &  1.14 &  1080 &  III  \\
1044 &  15.28 & k0.5 V      &  1.0 &  5.89 &  0.50 &   600 &  III  \\
1047 &  14.74 & f8 V        &  0.9 &  3.90 &  1.03 &   910 &  II  \\
1054 &  14.68 & f9 V        &  0.8 &  4.19 &  0.88 &   840 &  III \\
1055 &  13.96 & g0 V        &  1.0 &  4.48 &  0.32 &   680 &  II  \\
1056 &  14.68 & k4.5 V      &  1.0 &  7.08 &  0.50 &   264 &  III \\
1059 &  16.12 & g3 V        &  1.0 &  4.84 &  0.86 &  1210 &  II  \\
1061 &  12.45 & a8 V        &  0.9 &  2.45 &  0.74 &   710 &  III \\
1067 &  12.63 & f0 III      &  0.9 &  1.37 &  1.06 &  1100 &  II  \\
1070 &  13.93 & g1.5 V      &  1.0 &  4.51 &  0.45 &   620 &  III \\
1071 &  11.32 & k2.2 III    &  0.8 &  1.40 &  0.25 &   860 &  II  \\
1074 &  14.29 & k2.2 III    &  0.8 &  0.96 &  1.47 &  2360 &  II  \\
1076 &  10.00 & k0.5 III    &  1.0 &  0.52 &  0.42 &   650 &  II  \\
1077 &  14.14 & g8 III      &  1.0 &  1.10 &  1.44 &  2090 &  II  \\
1078 &  15.90 & k2 V        &  0.7 &  6.31 &  0.48 &   660 &  II  \\
1081 &  13.57 & g2 V        &  0.8 &  4.55 &  0.39 &   530 &  II  \\
1087 &  15.58 & k0 V        &  0.7 &  5.79 &  0.64 &   680 &  III \\
1089 &  15.04 & f5 V        &  0.9 &  3.45 &  1.01 &  1310 &  II  \\
1090 &  13.43 & g5 V        &  1.0 &  5.03 &  0.43 &   393 &  II  \\
1093 &  14.05 & g5 V        &  1.0 &  4.95 &  0.51 &   520 &  III \\
1096 &  14.54 & k1 III      &  0.8 &  1.02 &  1.72 &  2290 &  III \\
1101 &  15.68 & k1.2 V      &  0.8 &  6.12 &  0.40 &   680 &  II  \\
1102 &  11.35 & g9.5 III    &  1.0 &  0.88 &  0.79 &   860 &  II  \\
1105 &  15.83 & g8 V        &  1.0 &  5.00 &  0.62 &  1100 &  II  \\
1106 &  15.04 & g5 IV       &  1.0 &  2.72 &  1.86 &  1240 &  III  \\
1111 &  15.29 & g9 V        &  1.0 &  5.79 &  0.44 &   650 &  III  \\
1113 &   9.14 & k0 IV       &  1.0 &  2.78 &  0.30 &   164 &  II  \\
1114 &  13.87 & f9 V        &  0.9 &  4.14 &  0.40 &   730 &  III  \\
1116 &  14.54 & g1 V        &  0.9 &  4.37 &  0.98 &   690 &  II  \\
1117 &  14.68 & k1.5 V      &  1.0 &  6.16 &  0.45 &   412 &  III  \\
1120 &  13.02 & g1 IV       &  0.9 &  2.82 &  0.79 &   760 &  III  \\
1122 &  15.74 & g4 V        &  1.0 &  4.74 &  1.03 &   990 &  II  \\
1123 &  15.39 & g9.5 IV     &  0.8 &  3.47 &  1.60 &  1160 &  II  \\
1125 &  14.29 & f1 IV       &  0.9 &  2.09 &  1.34 &  1490 &  II  \\
1130 &  12.20 & a8 V        &  0.9 &  2.25 &  0.74 &   700 &  III  \\
1131 &  14.56 & k0.5 V      &  1.0 &  5.72 &  0.37 &   494 &  II  \\
1134 &  13.54 & f6 V        &  0.8 &  3.97 &  0.39 &   690 &  II  \\
1135 &  13.40 & k1 III      &  1.0 &  0.58 &  1.66 &  1700 &  II  \\
1140 &  12.75 & k3 III      &  1.0 &  0.27 &  1.91 &  1300 &  III  \\
1141 &  12.86 & g3 V        &  0.8 &  4.80 &  0.25 &   364 &  III  \\
1144 &  14.69 & k1 V        &  1.0 &  5.91 &  0.57 &   437 &  III  \\
1145 &  15.67 & g6 V        &  1.0 &  5.01 &  0.77 &   950 &  II  \\
1148 &  16.01 & k3 V        &  0.7 &  6.69 &  0.43 &   600 &  II  \\
1149 &  11.90 & g3 V        &  1.0 &  4.92 &  0.12 &   236 &  III  \\
1152 &  15.14 & k8 V        &  0.8 &  7.49 &  0.00 &   339 &  III  \\
1153 &   8.09 & f0 V        &  0.9 &  2.63 &  0.09 &   118 &  II  \\
1155 &  14.88 & k3.2 V      &  0.8 &  6.72 &  0.41 &   355 &  II  \\
1157 &  11.76 & a9 IV       &  0.9 &  2.00 &  0.63 &   670 &  II  \\
1158 &  14.81 & k0 IV       &  1.0 &  3.05 &  1.41 &  1170 &  II  \\
1161 &  13.76 & f9 V        &  0.8 &  4.26 &  0.30 &   690 &  II  \\
1162 &  12.28 & g0 V        &  0.8 &  4.20 &  0.31 &   359 &  II  \\
1163 &  14.44 & k0 V        &  1.0 &  5.91 &  0.25 &   453 &  III  \\
1164 &  14.99 & k0.7 V      &  1.0 &  5.66 &  0.37 &   620 &  II  \\
1166 &  15.76 & f3 V        &  0.8 &  3.16 &  1.54 &  1630 &  II  \\
1168 &  16.20 & f1 V        &  0.8 &  2.95 &  1.65 &  2090 &  II  \\
1170 &  15.39 & k3 V        &  0.8 &  6.64 &  0.27 &   497 &  II  \\
1173 &  12.62 & f3 V        &  0.9 &  3.16 &  0.44 &   640 &  II  \\
1175 &  13.24 & g8.5 IV     &  1.0 &  3.66 &  0.43 &   670 &  II  \\
1177 &  15.09 & g2 V        &  0.8 &  4.67 &  1.15 &   710 &  III  \\
1178 &  14.79 & f6 V        &  0.8 &  3.92 &  0.86 &  1000 &  II  \\
1180 &  14.45 & f9.5 IV     &  0.9 &  2.18 &  1.08 &  1730 &  II  \\
1183 &  15.91 & g1 V        &  0.8 &  4.58 &  0.84 &  1250 &  II  \\
1186 &  14.58 & g5 V        &  1.0 &  4.84 &  0.60 &   670 &  II  \\
1187 &  14.55 & g2 V        &  0.9 &  4.71 &  0.41 &   770 &  III  \\
1188 &  14.58 & g1.5 V      &  1.0 &  4.51 &  0.31 &   900 &  II  \\
1190 &  14.93 & f6 IV       &  0.9 &  2.43 &  1.56 &  1540 &  III   \\
1194 &  15.76 & k2.5 V      &  0.8 &  6.49 &  0.32 &   610 &  II  \\
1202 &  13.91 & g0 V        &  1.0 &  4.30 &  0.79 &   580 &  II  \\
1203 &  13.74 & g0 V        &  0.9 &  4.28 &  0.38 &   650 &  II  \\
1205 &  15.02 & f8 V        &  0.8 &  4.15 &  1.38 &   790 &  II  \\
1209 &  15.43 & k0 V        &  1.0 &  5.90 &  0.68 &   590 &  III  \\
1214 &  12.51 & k0.5 III    &  1.0 &  1.06 &  1.34 &  1050 &  II  \\
1215 &  13.29 & g9.5 III    &  1.0 &  0.75 &  1.46 &  1650 &  II  \\
1217 &  14.63 & g9.5 V      &  1.0 &  5.49 &  0.29 &   590 &  II  \\
1218 &  15.95 & f9 V        &  0.9 &  4.07 &  1.28 &  1320 &  II  \\
1219 &  16.01 & k3 V        &  0.8 &  6.53 &  0.40 &   660 &  II  \\
1221 &  10.54 & f0 IV       &  0.9 &  2.14 &  0.28 &   421 &  II  \\
1222 &  14.49 & g0 V        &  0.9 &  4.21 &  0.39 &   950 &  II  \\
1224 &  15.94 & k3 V        &  0.8 &  6.50 &  0.47 &   620 &  II  \\
1225 &  15.32 & g2 V        &  0.8 &  4.67 &  0.62 &  1010 &  II  \\
1226 &  12.63 & g1 V        &  0.9 &  4.31 &  0.23 &   414 &  II  \\
1227 &  13.32 & g5.5 V      &  1.0 &  5.10 &  0.37 &   372 &  II  \\
1228 &  14.31 & g7 V        &  1.0 &  5.23 &  0.65 &   486 &  II  \\
1233 &  14.78 & f9.5 V      &  0.8 &  4.39 &  0.29 &  1050 &  II  \\
1236 &  12.95 & g1.5 V      &  1.0 &  4.42 &  0.37 &   427 &  II  \\
1238 &  15.84 & k3.5 V      &  0.7 &  6.89 &  0.51 &   488 &  II  \\
1239 &  12.99 & f8 V        &  0.9 &  3.95 &  0.29 &   560 &  II  \\
1240 &  13.63 & g1 V        &  0.8 &  4.56 &  0.22 &   590 &  II  \\
1243 &  13.67 & g3 V        &  1.0 &  4.70 &  0.31 &   540 &  II  \\
\hline
\end{longtable}
}
\vskip4mm

\vbox{
\centerline{\psfig{figure=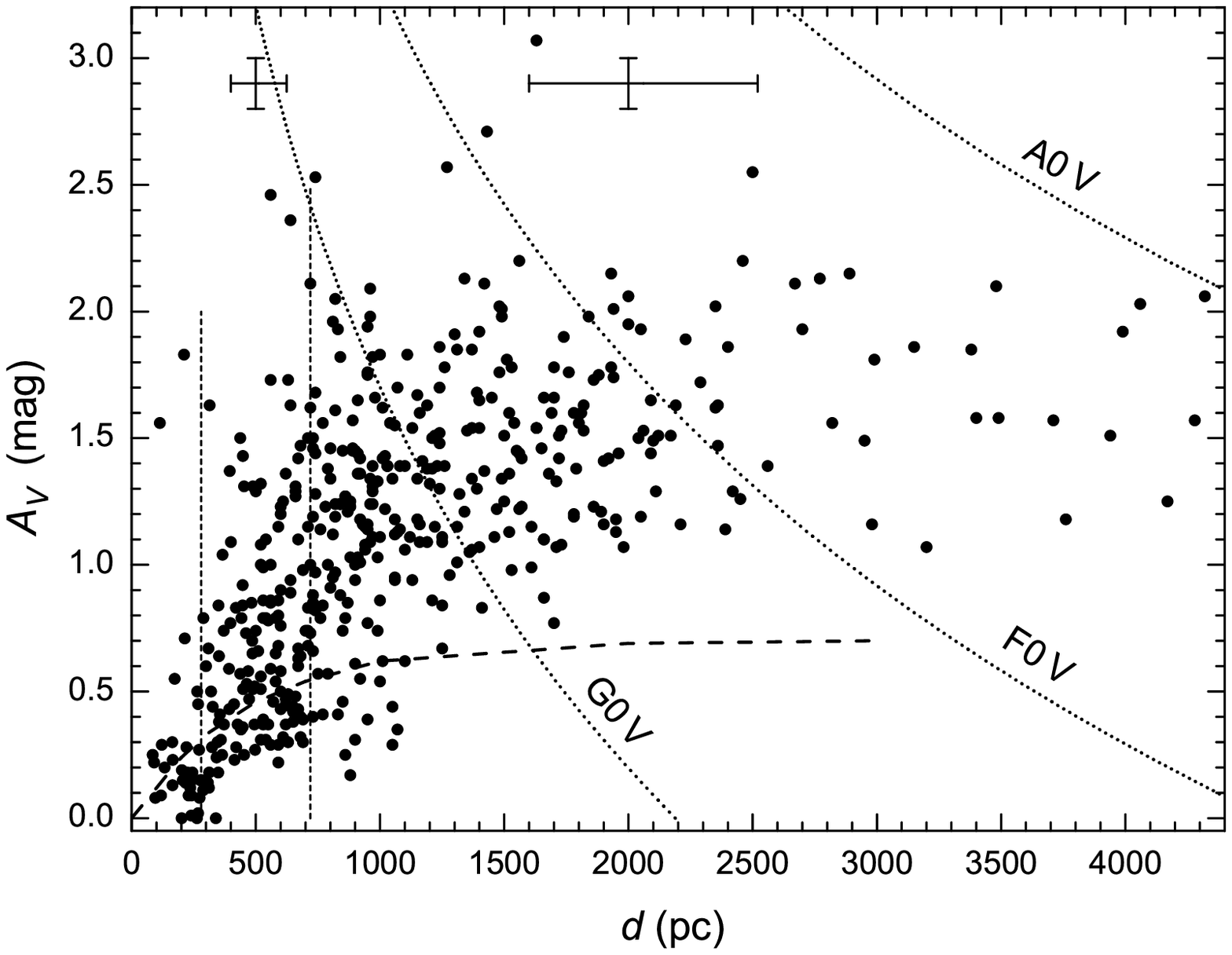,width=124mm,angle=0,clip=}}
\captionb{1}{Dependence of the $A_V$ extinction on distance in the whole
area.  The three dotted curves show the limiting magnitude effect for
A0\,V, F0\,V and G0\,V stars.  The curve at the right-hand upper corner
is also valid for K0\,III stars since their absolute magnitudes are
close to those of A0\,V stars.  The lower segmented curve is the
dependence of the extinction on distance for the Galactic latitude
+14.2\degr\ calculated by the Parenago formula (see the text).  The
error bars correspond to standard deviations of the distance and the
extinction at 0.5 kpc and 2 kpc distances. The two vertical lines
mark the mean distances of the clouds.}
\vskip2mm
}


\newpage

Figure 1 shows the plot $A_V$ vs.\,$d$ for stars in the whole area.  The
three dotted curves correspond to A0\,V (or K0\,III), F0\,V and G0\,V
stars at the limiting magnitude $V_{\rm lim}$ = 16.0.  The stars of
these spectral types (and absolutely fainter) above the corresponding
curves are affected by the limiting magnitude, i.e., stars with high
extinctions are missing at these distances.  Consequently, the plot
cannot be used for estimating both the mean and the maximum
extinctions.  However, up to a distance of 1 kpc all the stars
absolutely brighter than G0\,V are well represented in the areas
where $A_V$ is smaller than $\sim$\,2 mag.  At $d$ = 1 kpc and $A_V$ = 2
mag, only G, K and M dwarfs near the limiting magnitude are missing.  In
the upper part of Figure 1 the error bars of the distance and $A_V$ are
shown for the two distance values.  They correspond to an error of
$\pm$\,0.1 mag in $A_V$, $\pm$\,0.5 mag in $M_V$ and (--20, +26)\,\% in
the distance.

\vskip4mm
\vbox{
\centerline{\psfig{figure=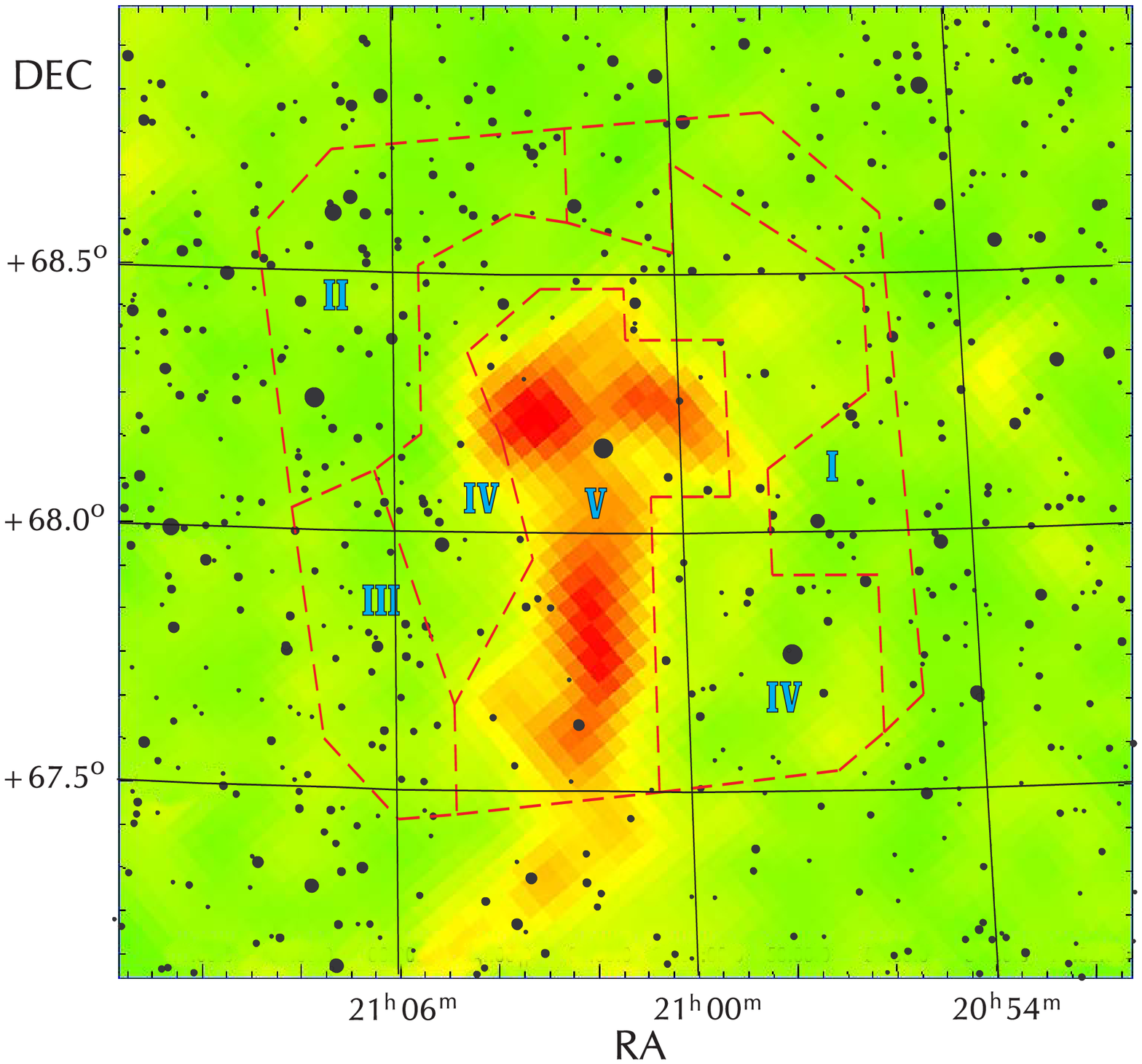,width=124mm,angle=0,clip=}}
\captionb{2}{The division of the investigated area into five subareas exhibiting
slightly different dependencies of $A_V$ on distance. The extinction map
from Dobashi et al. (2005) atlas and the stars down to $V$\,=\,14 from
GSC are shown in the background.}
\vskip4mm
}
\newpage

The segmented curve in Figure 1, which starts from the origin of the
coordinates, corresponds to the exponential extinction law for the
Galactic latitude $b$\,=\,+14.2\degr, calculated by the Parenago formula
with the extinction coefficient $A_V$\,=\,1.5 mag/kpc and the
half-thickness of the dust layer $\beta$ = 0.11 kpc (Parenago 1945;
Sharov 1963; Strai\v{z}ys 1992, p.\,146).  It is evident that the
Parenago curve is in agreement with the positions of low-extinction
stars located closer to us than 500--700 pc.

For determining the distance to a dark cloud we usually use stars
situated at a steep rise (or jump) of the extinction at the front edge
of the cloud.  However, some of these stars can have negative distance
errors which originate mainly from the errors in their absolute
magnitudes.  Consequently, the true distance to the cloud can be larger
than the distance corresponding to the jump defined by the stars
apparently closest to the Sun.  The true distance can be found as $d$ =
$d$\,(front) + 0.2\,$d$, or $d$ = $d$\,(front)\,/\,0.8, where 0.2\,$d$
is the negative distance error when the error of the absolute magnitude
$\Delta M_V$ = +0.5.

However, the described situation takes place only in the case when a
statistically significant number of stars at the extinction jump is
available.  In our case, at the expected cloud distance we have only a
few stars with large extinctions.  It is quite possible that some of
them really have a negative error of the distance but it is also
possible that their distance error happens to be zero or even positive,
and we have no reason to apply the above described correction to their
apparent distances.  A more realistic value of the cloud distance can be
obtained by averaging distances of the reddened stars in the interval
between $d-0.20d$ and $d+0.26d$ where $d$ is the true cloud distance.

In Figure 1 we can see that at $\sim$\,250 pc a steep rise in the
extinction takes place.  However, two of the stars, Nos.\,595 and 636 in
the catalog of Paper I, exhibit too large extinction values, $A_V$ =
1.56 and 0.55 mag, at small distances, 114 pc and 173 pc, respectively.
The first of these two stars will be discussed below in this section.
Both stars are excluded from determining the cloud distance.  The
remaining 10 stars with distances between 210 and 330 pc and
$A_V$\,$\geq$\,0.5 mag have the mean distance 282\,$\pm$\,42 pc
(standard deviation) which may be considered as the distance of the
nearest cloud.  This result should be considered as more accurate than
the value of distance found by Strai\v{z}ys et al.  (1992) applying a
similar method for only four stars of magnitudes 11--12.  Two of them in
our Paper I were suspected as binaries.

At distances larger than 250 pc the extinction continues to rise almost
up to 1 kpc.  However, the presence of another jump (or jumps) of the
extinction can be suspected.  The most probable jump is observed between
560 and 875 pc, where 560 pc is the distance to the front edge, and the
distance range is defined by $d-0.20d$ and $d+0.26d$.  Within this
distance range we have 10 stars with $A_V$\,$\geq$\,1.8 mag.  Their mean
distance is 715\,$\pm$\,110 pc (standard deviation).

Trying to better understand the extinction vs. distance relation, we
have split the investigated field into five subareas with the boundaries
shown in Figure 2 and with the extinction map from the Dobashi et al.
(2005) atlas and the stars down $V$\,=\,14 mag from the GSC catalog
plotted in the background.  Each of these subareas exhibits a somewhat
different form of the $A_V$ vs. distance dependence.  In the following,
the results of the extinction dependence on distance in these subareas
will be described.


\vbox{
\centerline{\psfig{figure=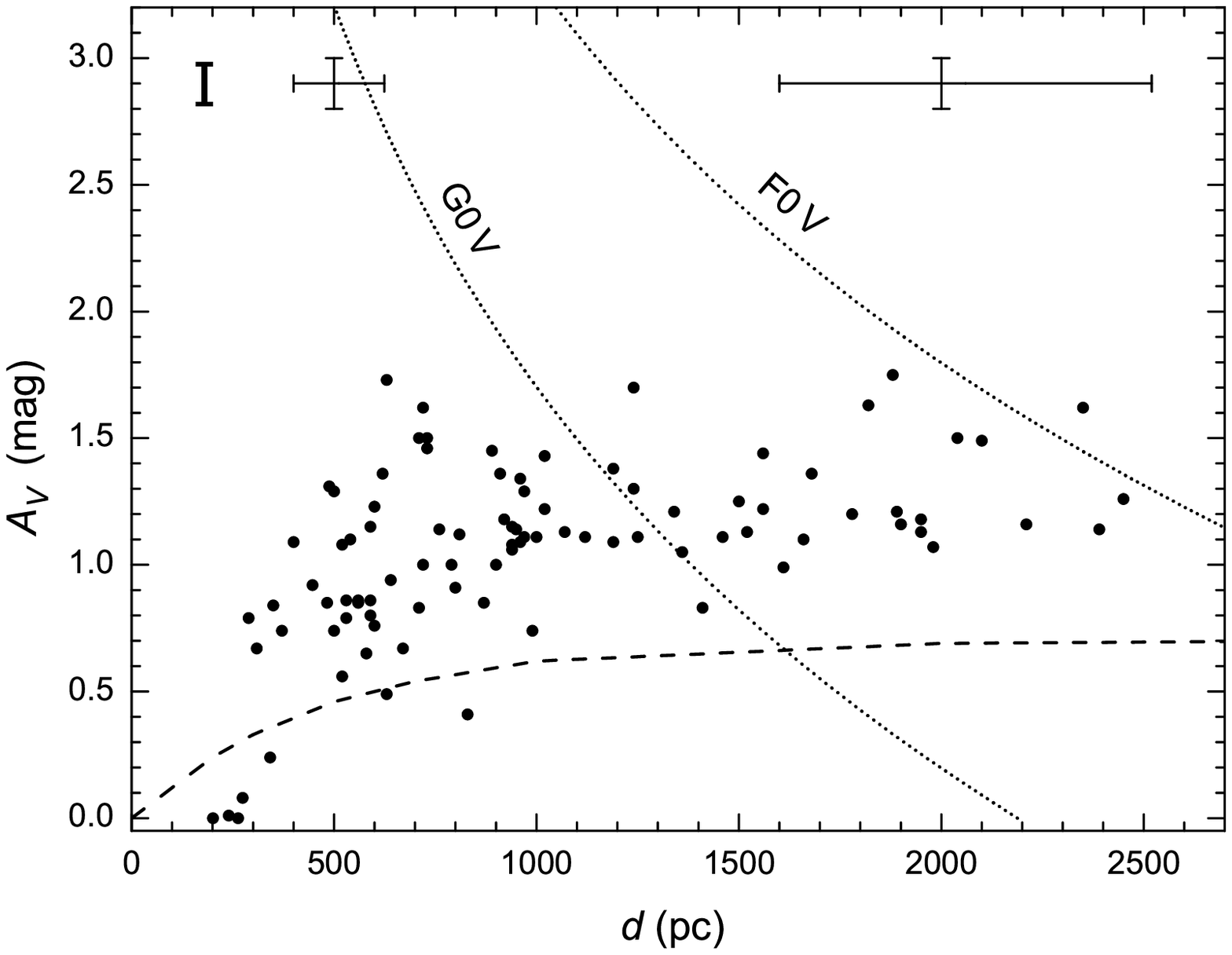,width=110mm,angle=0,clip=}}
\vspace{-5mm}
\captionc{3}{The same as in Figure 1 but for Subarea I.}
\vskip7mm
\centerline{\psfig{figure=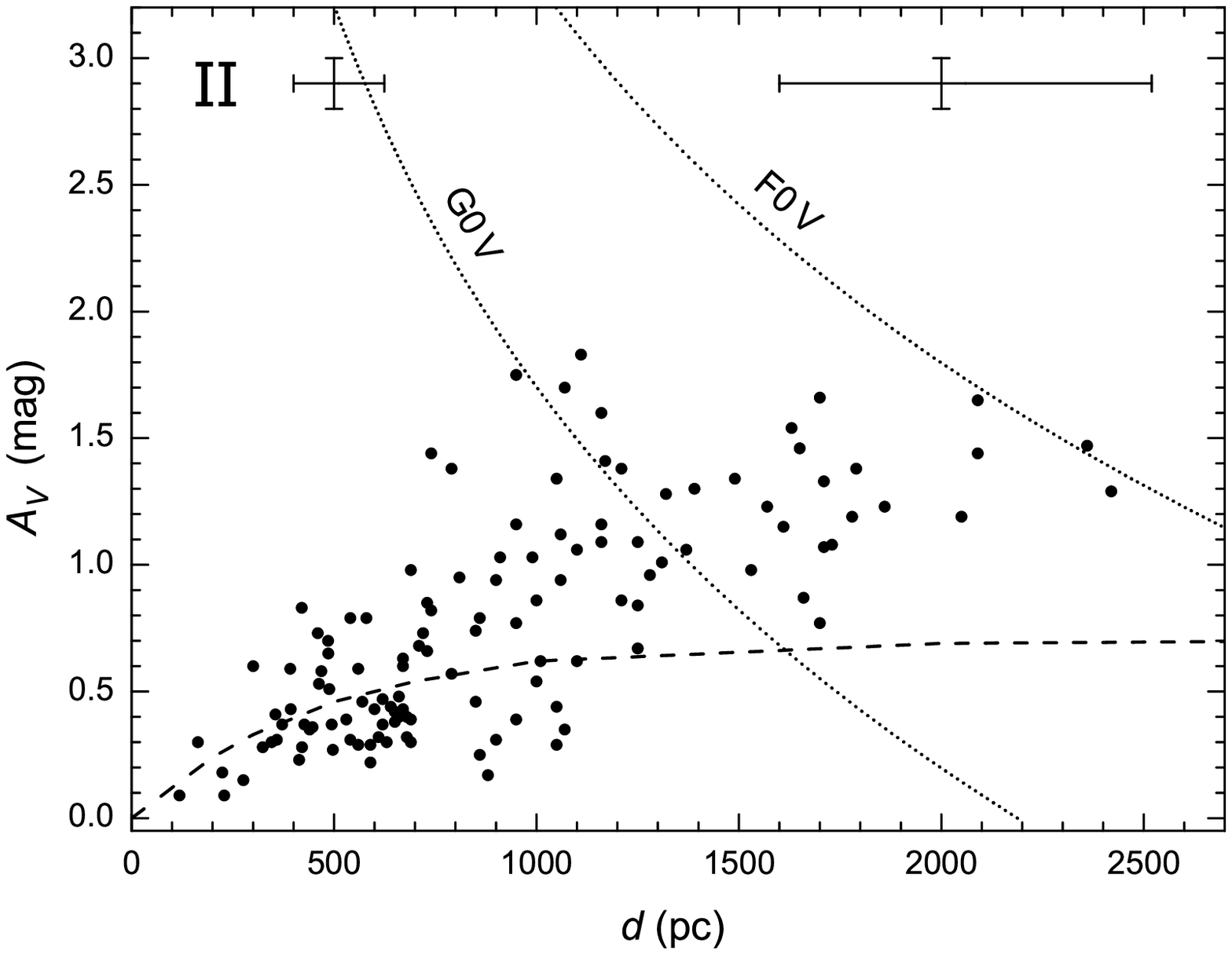,width=110mm,angle=0,clip=}}
\vspace{-5mm}
\captionc{4}{The same as in Figure 1 but for Subarea II.}
}

\vbox{
\centerline{\psfig{figure=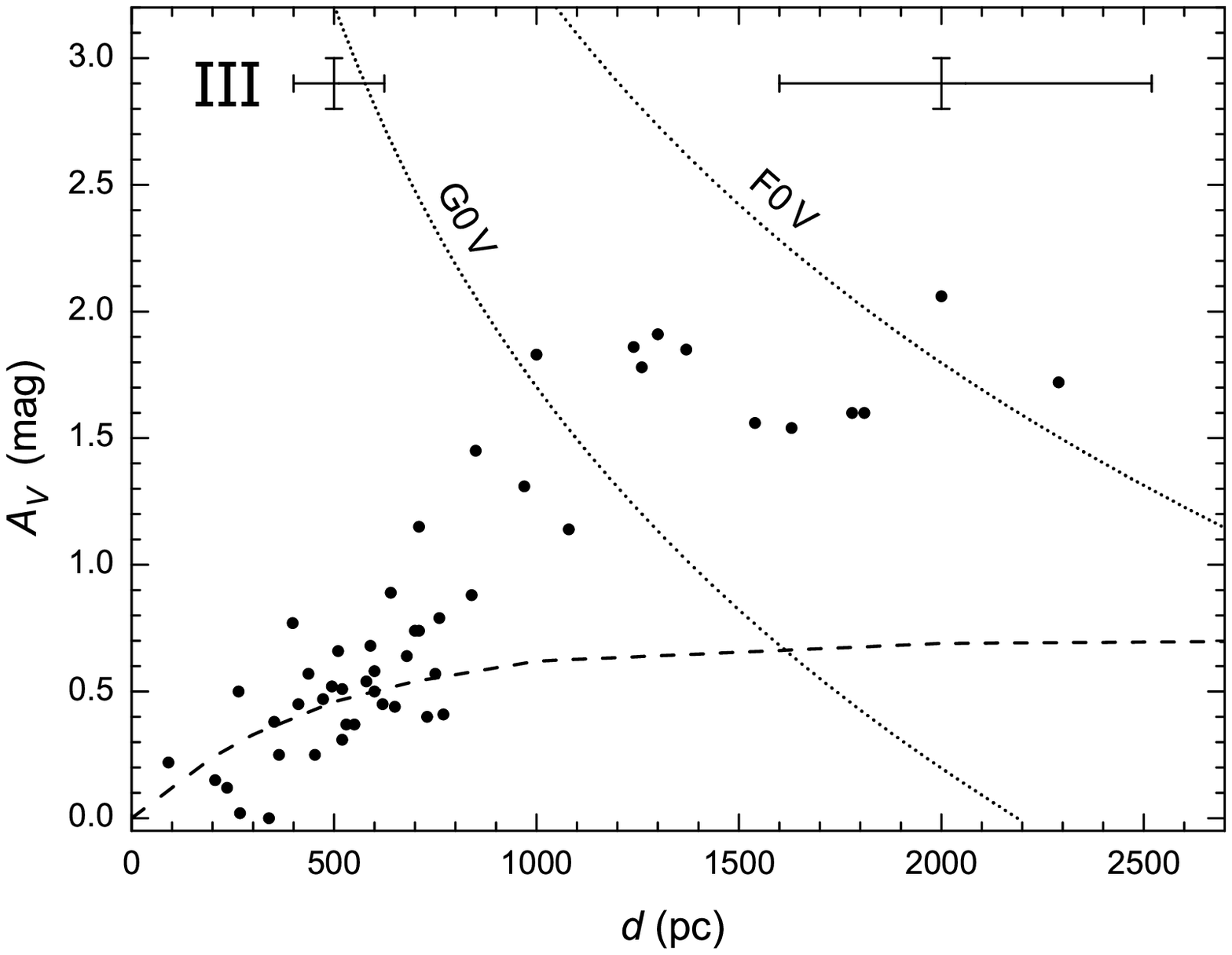,width=110mm,angle=0,clip=}}
\vspace{-5mm}
\captionc{5}{The same as in Figure 1 but for Subarea III.}
\vskip7mm
\centerline{\psfig{figure=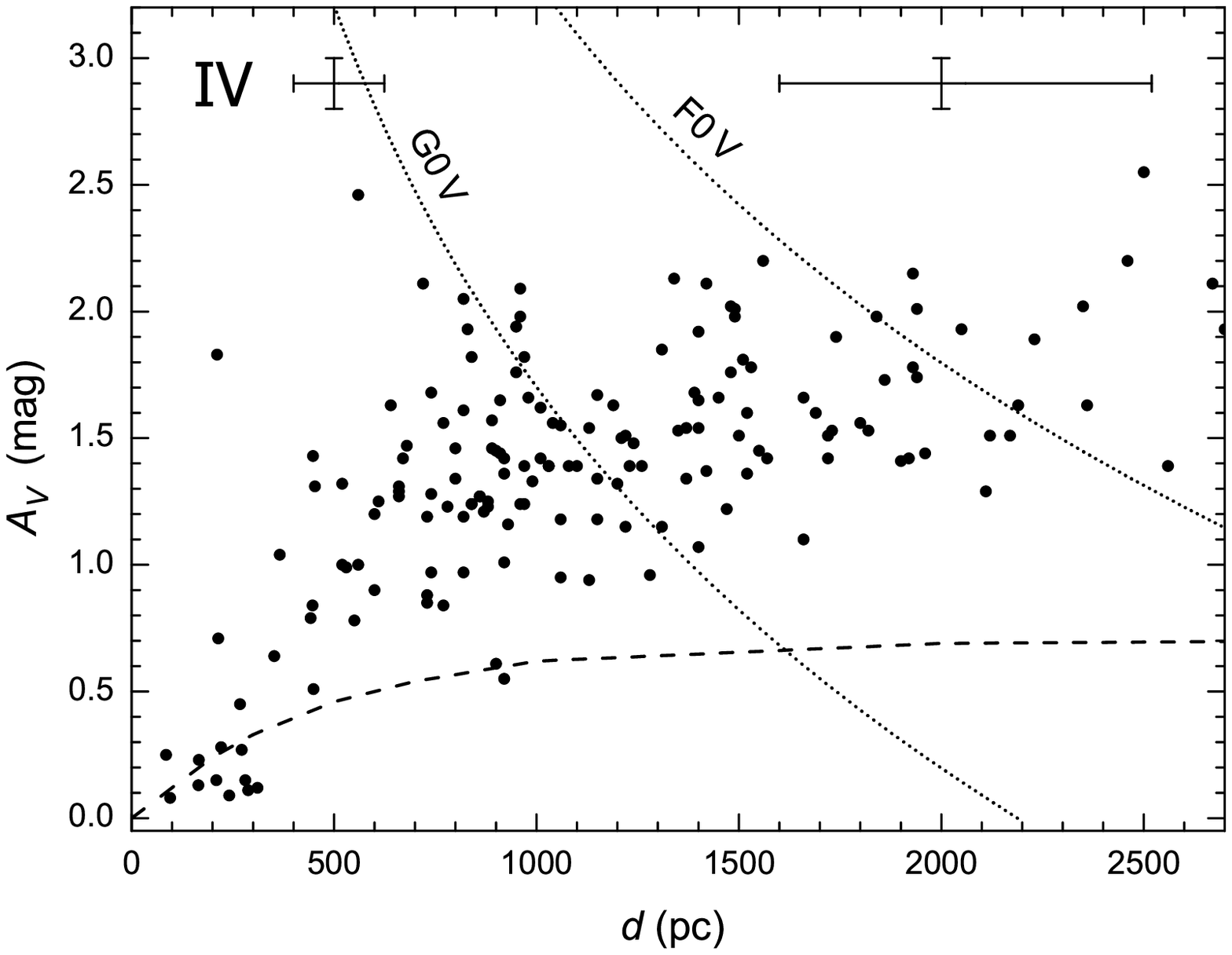,width=110mm,angle=0,clip=}}
\vspace{-5mm}
\captionc{6}{The same as in Figure 1 but for Subarea IV.}
}

\vbox{
\centerline{\psfig{figure=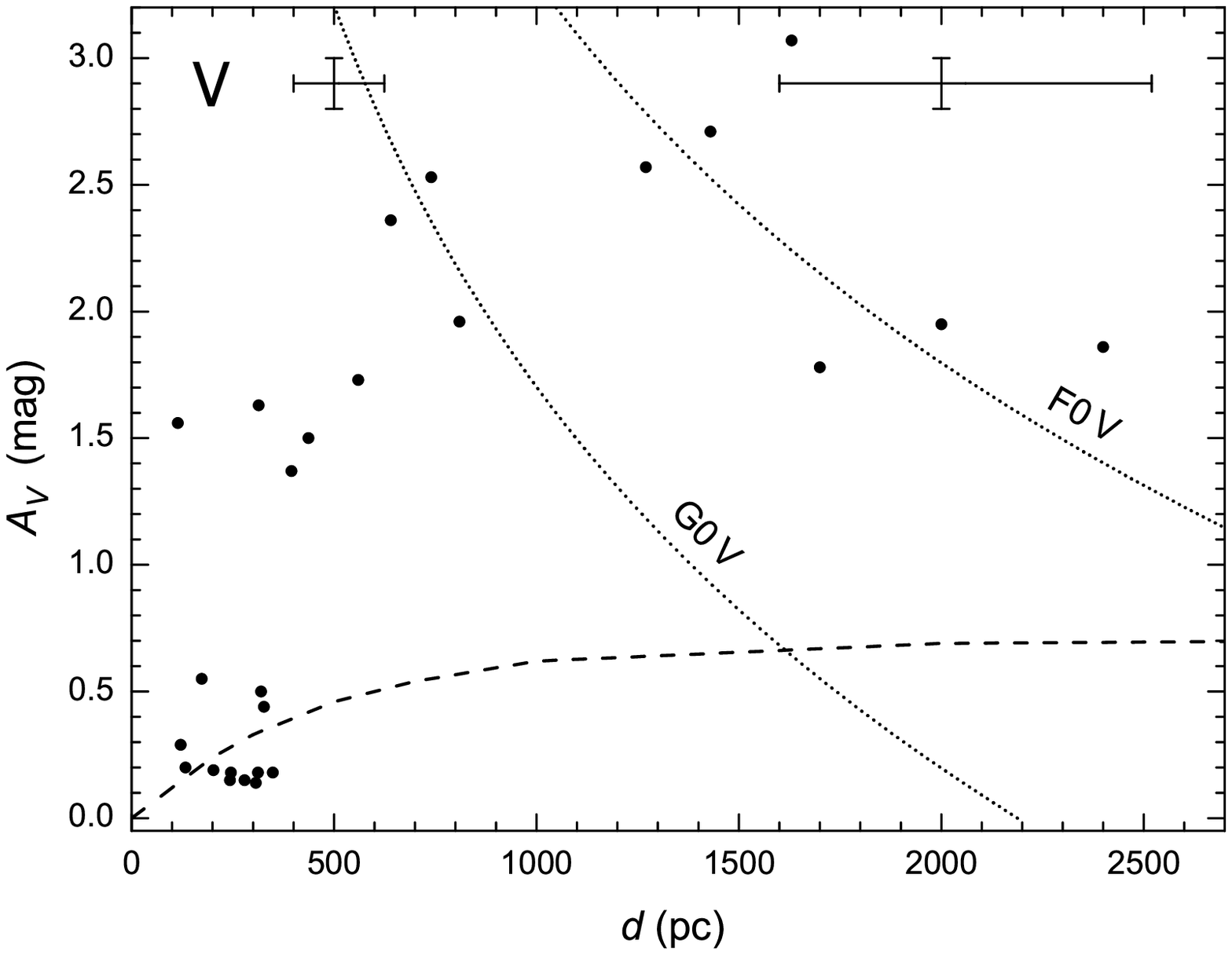,width=110mm,angle=0,clip=}}
\vspace{-5mm}
\captionc{7}{The same as in Figure 1 but for Subarea V.}
\vspace{8mm}
}

Figure 3 shows the $A_V$ vs.\,$d$ plot for Subarea I located along the
right edge of the field.  The first two reddened stars with $A_V$
between 0.6--0.8 mag are seen at an apparent distance of $\sim$\,290 pc,
i.e., quite close to the mean distance of the first cloud estimated from
Figure 1. A few more jumps between 500 pc and 750 pc are also possible.
The mean extinction value at distances $>$\,1.0 kpc is about 1.3 mag,
and the maximum value is close to 1.75 mag.

Figure 4 shows the $A_V$ vs.\,$d$ plot for Subarea II located at the
left upper corner of the area.  Here, the nearest considerably reddened
star is found at the apparent distance 300 pc, and the second jump is
seen at $\sim$\,700 pc.  At $d$\,$>$\,1 kpc the extinction remains more
or less constant with a mean value of 1.3 mag.  In this subarea a group
of about 12 stars at a distance of 800--1100 pc exhibits quite low
extinction, with the values between 0.2 and 0.6 mag.  Probably, these
stars are seen in the directions of relatively transparent windows.
They are scattered over the whole subarea.

Figure 5 shows the $A_V$ vs.\,$d$ plot for Subarea III located at the
lower left corner of the field.  The positions of the two extinction
jumps here cannot be estimated reliably but the height of the second
jump is almost 1 mag, giving a mean extinction of 1.8 mag at
$d$\,$>$\,1~kpc.

Figure 6 shows the $A_V$ vs.\,$d$ plot for Subarea IV which surrounds
the central dark cloud on three sides.  The extinction jumps are close
to the distances observed in other subareas.  The extinction values show
a considerable scatter (between 1.0 and 2.2 mag), with the mean value
being about 1.6 mag.  Two stars in the subarea exhibit the extinction
values around 2.5 mag.

In Figure 7 we show the $A_V$ vs.\,$d$ plot for Subarea V which includes
the darkest segment of the dust cloud with the reflection nebula
NGC\,7023.  Only 14 classified stars with $A_V$\,$>$\,1.0 have been
found in this direction.  Among these the most interesting is the
above-mentioned star No.\,595.  Its photometric spectral type is K2\,V,
$V$ = 13.18, $A_V$ = 1.56 and $d$ = 114 pc.  It is strange to find this
large extinction at such a small distance.  The classification of the
star by all of the methods applied is of good accuracy and coinciding.
The small apparent distance of the star can be explained by its possible
duplicity.  If it is a binary with two identical components, the
combined absolute magnitude should be more negative by 0.75 mag and the
distance larger by a factor of 1.41, i.e., 114\,$\times$\,1.41 = 161 pc,
which is more realistic than the value for a single star, but still too
small compared to the cloud distances in other subareas.

Other stars in Subarea V classified in Paper I are too scanty to
estimate cloud distances.  However, their distribution in the plot
(Figure 7) is not in contradiction to the apparent distances of the two
clouds at 282 pc and 715 pc.  The largest extinction found in Subarea V
is close to 3 mag, but this is not the real maximum value since the
stars with larger extinctions are absent in our sample due to the
limiting magnitude effect.  With the help of 2MASS photometry we have
found in this area a few red giants having $A_V$\,$\approx$\,15 mag.

\sectionb{4}{DISCUSSION AND CONCLUSIONS}

The dust cloud TGU\,629, surrounding the reflection nebula NGC\,7023,
belongs to a giant dust and molecular cloud system known as the Cepheus
Flare.  In the summaries of distance determinations of different objects
in this system, Kun (1998) and Kun et al.  (2008) came to the conclusion
that the system either has a considerable depth or consists of several
layers with distances ranging from 200 to 500 pc.  Two layers of
interstellar gas were found by radio observations by Heiles (1967) in
the neutral hydrogen 21 cm line and by Grenier et al.  (1989) in the CO
molecular lines.  Applying the kinematical method to velocity profiles
of the lines, Grenier et al. find the approximate distances to the
layers:  300 and 800--900 pc.

Our results described in Section 3 also give evidence that dust clouds
in the vicinity of NGC\,7023 concentrate at least in two layers at 282
pc and 715 pc.  There is a possibility that the true distances of these
cloud layers are not the same throughout the area.  However, the number
of stars at the extinction jumps in different subareas is too small to
be sure that these distance differences are real.  The extinction vs.
distance plots also allow to suspect that more clouds are present
along the line of sight.  This is in agreement with the map of the
CO intensity distribution (Dame et al. 2001) which evidences that the
molecular cloud structure in the Cepheus Flare is quite clumpy and
fragmented.

Our estimates of cloud distances are in satisfactory agreement with
those found by Grenier et al.  (1989) from kinematics of the CO clouds.
The CO radial velocities show that at the Galactic longitude
$\ell$\,=\,104\degr\ both clouds are connected by a bridge.  The distant
CO layer should be more prominent at larger Galactic longitudes, i.e.,
on the left side of our area (Subareas II, III and, partly, IV).  This
is in agreement with our results.

If we accept that dust clouds in this direction reach a distance of 700
pc, the depth of the cloud layer should be about 400 pc.  The length of
the whole Cepheus Flare cloud system ($\sim$\,18\degr) corresponds to
$\sim$\,95 pc at a distance of 300 pc and to $\sim$\,220 pc at a
distance of 700 pc.  It seems possible that the Cepheus Flare has its
extension known as the Polaris Flare (Heithausen et al. 1993; Dame et
al. 2001).  In this case the whole complex of molecular clouds from
$\ell$, $b$ = (100\degr, +14\degr) to (126\degr, +30\degr) has a length
of $\sim$\,30\degr\ and the projected complex length is from $\sim$\,160
pc at a distance of 300 pc to $\sim$\,375 pc at a distance of 700 pc.
The apparent width of the Cepheus and Polaris Flares is only
$\sim$\,8\degr, which corresponds to 42 pc at a distance of 300 pc and
100 pc at 700 pc.

The projected length of the cloud system at 700 pc (375 pc) is
comparable to the observed depth of the complex (400 pc), i.e., the
complex looks like a pancake, and our line of sight runs along its
plane.  The heights of the two cloud layers above the Galactic plane in
the direction of NGC\,7023 are 75 pc and 170 pc.

To have the estimates of cloud distances more accurate, one must
minimize the errors of absolute magnitudes of the stars which define the
jumps in the extinction vs. distance dependence.  This can be done
either by spectral observations of these stars to verify their spectral
and luminosity classes or by determining trigonometric parallaxes.
Within a few years, in the case of the success of the {\it Gaia}
mission, the distance problem of these reddened stars will be solved.

\thanks{ We are thankful to Edmundas Mei\v{s}tas and Stani\-slava
Barta\v{s}iute for their help in preparing the paper.  The use of the
SkyView, Simbad, Gator and 2MASS databases is acknowledged.}

\References

\refb Cutri R. M., Skrutskie M. F., Van Dyk S., Beichman C. A. et al.
2003, {\it 2MASS All Sky Catalog of Point Sources}, NASA/IPAC Infrared
Science Archive,\\ http://irsa.ipac.caltech.edu/applications/Gator/

\refb Dame T. M., Hartmann D., Thaddeus P. 2001, ApJ, 547, 792

\refb Dobashi K., Uehara H., Kandori R., Sakurai T., Kaiden M.,
Umemoto T.,\\ Sato F. 2005, PASJ, 57, S1

\refb Grenier I. A., Lebrun F., Arnaud M., Dame T. M., Thaddeus P. 1989,
ApJ, 347, 231

\refb Heiles C. 1967, ApJS, 15, 97

\refb Heithausen A., Stacy J. G., de Vries H. W., Mebold U., Thaddeus P.
1993, A\&A, 268, 265

\refb Kun M. 1998, ApJS, 115, 59

\refb Kun M., Kiss Z. T., Balog Z. 2008, in {\it Handbook of Star
Forming Regions},\\ ed. B. Reipurth, ASP, vol. 1, p.\,136

\refb Parenago P. P. 1945, AZh, 22, 129

\refb Sharov A. S. 1963, AZh, 40, 900 = Soviet Astron., 7, 689

\refb Skrutskie M. F., Cutri R. M., Stiening R., Weinberg M. D. et al.
2006, AJ, 131, 1163

\refb Strai\v{z}ys V. 1992, {\it Multicolor Stellar Photometry},
Pachart Publishing  House,\\ Tucson, Arizona

\refb Strai\v{z}ys V., \v{C}ernis K., Kazlauskas A., Laugalys V. 2002,
Baltic Astronomy, 11, 219

\refb Strai\v{z}ys V., \v{C}ernis K., Kazlauskas A., Mei\v{s}tas E.
1992, Baltic  Astronomy, 1, 149

\refb Zdanavi\v{c}ius K. 2005, Baltic Astronomy, 14, 104

\refb Zdanavi\v{c}ius K., Zdanavi\v{c}ius J., Strai\v{z}ys V., Kotovas
A. 2008, Baltic Astronomy, 17, 161 (Paper I)

\end{document}